\documentclass[11pt,twoside]{article}
\usepackage{asp2010}

\resetcounters

\begin{document}

\title{Probing the formation of relativistic jets in real-time with SPICA}
\author{P. Gandhi,$^1$ D.M. Russell,$^2$ P. Casella,$^3$ J. Malzac,$^{4}$ T. Nakagawa$^5$
\affil{$^1$Department of Physics, Durham University, UK}
\affil{$^2$Instituto de Astrof\'{i}sica de Canarias, Spain}
\affil{$^3$INAF - Osservatorio Astronomico di Roma, Rome, Italy}
\affil{$^4$Universit\'{e} de Toulouse and CNRS, Toulouse, France}
\affil{$^5$Institute of Space and Astronautical Science, JAXA, Japan}}

\begin{abstract}
The time domain remains, in many respects, the least explored of parameter spaces in astronomical studies. The purpose of this article is to encourage the SPICA community to consider the potential of rapid infrared timing observations. The specific example considered is that of variable emission from relativistic jets in compact accreting objects, whose formation and powering mechanisms we still do not understand. Infrared observations have the potential to give us fundamental insight on the conditions required for jet formation in accreting stellar-mass black holes. This is because particle acceleration is thought to be magnetically-driven, and the spectral transition between optically-thin and self-absorbed jet synchrotron radiation lies in the infrared. We review recent observations from WISE showing that we have the capability to measure key physical parameters of the jet, and their time-dependence on rapidly-changing conditions in the accretion flow around the black hole (on timescales of just a few seconds). SPICA will provide a breakthrough in this field because of its sensitivity and broadband coverage, and we detail an example SPICA observation on short (tens of milliseconds) timescales. We believe that SPICA has the potential to make great impact on time domain science, and we discuss some technical requirements that will enable this.
\end{abstract}

\section{Introduction}

Black holes are fundamentally simple objects. Only two parameters -- mass and spin -- are required for their unique characterization. Yet, they host a variety of complex and extreme astrophysical phenomena in the vicinity. Accretion of gas drives their growth and much of their observed radiative power. Outflows in the form of relativistic \lq jets\rq\ and winds remove angular momentum, and are responsible for mechanical and thermal feedback which is now known to play a key role in galaxy evolution. 

Jets have been observed to extend linearly over scales as large as $\sim$Mpc from black holes (BHs) of the supermassive variety. Yet, many basic questions remain unanswered. For instance, what conditions give rise to a jet in some BHs and not others? And what fraction of the accretion energy is carried away by the jet as radiatively-invisible feedback? It is believed that magnetic fields likely play a key role in jet formation \citep[see, e.g., articles in ][]{belloni10}. But in order to understand this role, it is first necessary to measure the field strengths ($B$) at the base of the jets where they are form above BHs.

\section{Jets in stellar mass black holes}

The scaling with mass of BH physics means that it is easier to study the response of jets to varying conditions in the accreting material in {\em stellar mass} BH and neutron star binaries (Fig.~1), rather than in their supermassive cousins.\footnote{e.g. a timescale of 1 s for a 10 M$_{\odot}$ BH is equivalent to several months for a 10$^8$ M$_{\odot}$ supermassive BH. The former is much more amenable to observations.}

According to the \lq standard\rq\ model of emission from compact jets \citep{blandford79}, a peak in the flux density of jet emission is expected at a frequency dependent on the $B$ field and inversely dependent on the size of the jet base ($R$). This peak occurs because of magnetic and particle flux conservation, leading to each region in the jet being charcterized by a frequency that decreases with distance from the base. The sum envelope of all components is an inverted spectrum ($\alpha \geq 0$, where $F_{\nu}\propto \nu^{\alpha}$) up to a peak frequency associated with the optically thick-to-thin break ($\nu_{\rm break}$). At higher frequencies, optically-thin synchrotron ($\alpha < 0$) tells us the jet particle distribution. Some of the best current constraints show that $\nu_{\rm break}$ in BH binaries can straddle the entire infrared regime and more \citep[e.g. ][]{migliari10, rahoui11, g11_wise, corbel13, russell13_breaks}. See, e.g. the broadband jet break in the BH binary GX 339--4 around 10 $\mu$m shown in Fig. 2. 

In addition, broadband detections of GX 339--4 over multiple epochs with the WISE mission \citep{wise} have shown that $\nu_{\rm break}$ can change dramatically (by factors of at least $\sim$10) in just a few hours, and probably much faster. In fact, \citep[][ Fig. 3]{g11_wise} find variability on the shortest WISE cycle time of 11 s. This means that either $B$ or $R$ must be undergoing similar vacillations, on timescales faster than have been considered before. We are likely seeing the inner jet responding to stochastic variability in the accretion flow that feeds it, giving direct \lq real-time\rq\ constraints on the disk--jet connection through mid-infrared observations.

Yet, if we ask what are the physically most interesting timescales for probing such a connection, these turn out to be on the order of fractions of a second (corresponding to the dynamical time in the inner disk or the light travel time from the disk to the expected base of the jet) for a typical 10~M$_{\odot}$ BH. There is plenty of evidence for X-ray variability on these timescales. Recent observations have also found optical and near-infrared variations on fast sub-second timescales (see Fig. 3), at least part of which is correlated with the X-rays \citep{kanbach01, g08, durant09, g10, casella10, durant11}. So the next step has to be the search for fast (sub-second) {\em mid-infrared} variations that are correlated with X-rays.

\begin{figure}[!ht]
\begin{center}
   \resizebox{0.6\hsize}{!}{
     \includegraphics*{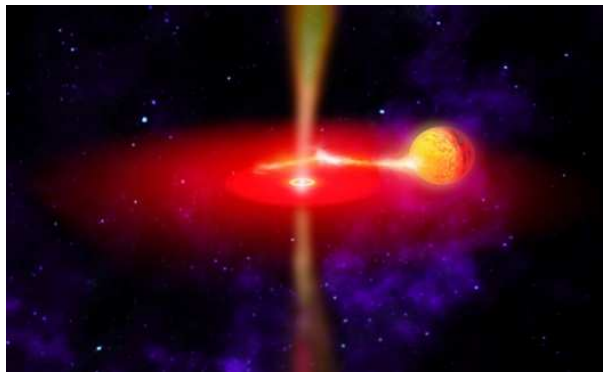}
   }
\end{center}
\caption{
Artist's illustration of an accreting stellar-mass black hole. The black hole accretes material from a companion donor star. Energy is extracted in the accretion disk and a fraction of the material is ejected in extended bipolar relativistic jets. Credit: NASA ({\tt www.nasa.gov/mission\_pages/WISE/news/wise20110920.html}). It is not possible to spatially resolve these components (except the extended jet in some binaries), so variability studies provide the best clues to the inferred structure. 
}\label{fig:xrb}
\end{figure}

\section{The need for SPICA}

The above results highlight the utility of the infrared regime for jet studies in stellar-mass BHs. However, most accreting stellar-mass binaries still remain undetected in the infrared. This is due to a combination of several reasons. Strong emission from accreting, stellar or surrounding material can dominate fainter jet radiation over the infrared--X-ray regime in many sources \citep[e.g. ][]{russell06}. In addition, the transient nature of activity in many binaries makes it difficult to catch them contemporaneously with multiple observatories necessary to probe broadband emission and isolate the jet. Finally, there has simply been a lack of monitoring instruments sensitive enough to detect infrared jet emission at sub-mJy flux levels, where most binaries lie. 

SPICA's infrared coverage over a broad infrared regime, as well as its superb sensitivity \citep{spica}, make it ideal for jet timing studies. 

\section{An example SPICA observation}

Here is a detailed example of the kind of observation to break new ground in this field. 

\begin{enumerate}
  \item One would wait for a Galactic binary to go into outburst, and trigger SPICA observations based upon some X-ray, radio or other multi-wavelength activity threshold. There are various monitoring missions (e.g. MAXI) well-suited for deciding when to trigger. 
  \item SPICA/MCS instrument \citep{spica_mcs} will be used in imaging or low-resolution mode, simultaneously with the near-infrared FPC \citep{spica_fpc}, covering $\sim$0.7--38 $\mu$m. The main aim is to cover the widest possible range of infrared wavelengths, {\em simultaneously}. Note that SAFARI \citep{spica_safari} may also be used separately after MCS observations. These will be non-simultaneous but still useful because of the broadband nature of jet emission.
  \item The simultaneous data will be used to locate $\nu_{\rm break}$ over the entire near- to mid-infrared (or more) regime covered, which, in turn, will determine physical parameters of the jet including $B$ and $R$ \citep[see ][ for details]{g11_wise}.
  \item SPICA has the potential to carry out such measurements on timescales of a fraction of a second. For instance, if the option of windowing and reading out only a small portion of the detector centered on the target can be implemented, this will enable continuous fast timing light curves to be measured. If detector readout time scales with pixel area, a $\sim$10$\times$10 pixel (1$\farcs$4$\times$1$\farcs$4) window on the MCS could yield an on-source time of $\sim$10 ms for the short wavelength arm, for instance. Extrapolating from the current expectation of the 1 hour imaging sensitivity assuming Poisson statistics scaling, the signal:noise achievable even on this short time will be sufficient for bright transient outbursts of some binaries (See Fig. 2). This assumes that the high zodiacal sky background dominates the noise. 
  \item The final pi{\`e}ce de r\'{e}sistance will be to coordinate the SPICA observation with strictly-simultaneous observations at another wavelength. For instance, an X-ray facility such as Astro-H \citep{astroh12} can be used to probe the accreting matter, while SPICA measures the response of the jet in real-time, enabling us to watch the interplay of accretion and outflows unfold before our (telescopic) eyes. 
\end{enumerate}

\section{Fast timing with SPICA: requirements and potential}

SPICA already possesses the most crucial requirements for the fast timing science described herein: the ability to observe simultaneously over a broad wavelength regime, and the requisite sensitivity. Below are some additional requirements, {\em none of which are particularly challenging if included in the initial mission planning stages now.} 

\begin{itemize}
  \item A provision should be made for Target of opportunity (ToO) observations. For the science described herein, a reaction time of up to a few days will be acceptable.
  \item The option of being able to window the detector will be very useful to increase the cycle time of exposures. Characterization of the \lq dead time\rq\ between consecutive exposures and any extra sources of noise on fast sampling times should also be carried out. Reaching down to $\sim$10 ms cycle times may be challenging, but if arbitrary window sizes and shapes can be coded in the detector control software, that would increase the versatility of the instrument significantly.
  \item Finally, coordination between different missions requires accurate clock calibrations for both {\em relative} and {\em absolute} timing. X-ray missions routinely achieve timing accuracies of fractions of a millisecond. Calibration of the on-board clock should not be an overly demanding task if this is planned for in advance.
\end{itemize}

\noindent
Such provisions will extend SPICA's reach into the time domain, allowing science ranging from exoplanet transits, to asteroseismology, pulsars and blazars, to name a few.

\section{Summary}

SPICA has huge potential for the study of transient phenomena, expanding its reach into the {\em hot}, non-thermal universe. The infrared regime is particularly important for the science of relativistic jets described herein, but a little bit of planning will allow SPICA to make its mark on time domain studies in general.

\begin{figure}[!ht]
\begin{center}
   \resizebox{0.99\hsize}{!}{
     \includegraphics*[angle=90]{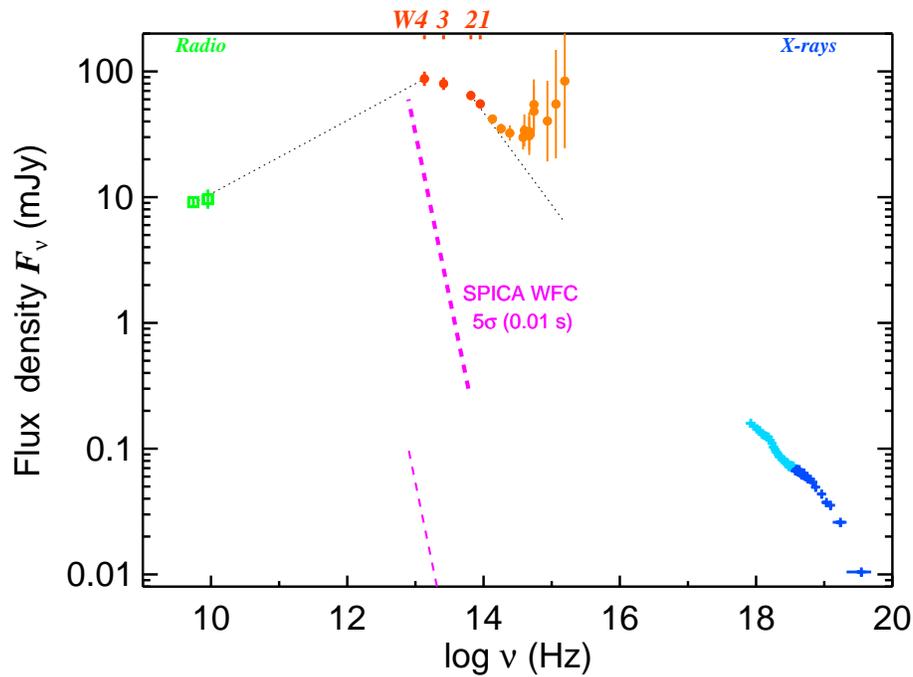}
   }
\end{center}
\caption{
Broadband quasi-simultaneous spectral energy distribution (SED) of the BH binary GX 339--4 during its 2010 active state \citep{g11_wise}. The mid-infrared peak probed by WISE is where the jet synchrotron spectral break ($\nu_{\rm break}$) lies. The thick dashed line shows the approximate 5$\sigma$ sensitivity limit of the MCS/WFC in only 10 ms (extrapolated from {\tt www.ir.isas.jaxa.jp/SPICA/spica2013/img/SPICA\_MCS\_130522a.pdf}, assuming Poisson noise scaling). This 10 ms sensitivity level is roughly comparable to the present WISE sensitivity with exposure times of about 8--9 s, over a wider wavelength regime than WISE. Continuous fast timing with SPICA on these timescales will allow us to probe entirely new parameter space. For comparison, the approximate 1 hour 5$\sigma$ sensitivity level is shown by the lower thin, dashed line.
}\label{fig:sed}
\end{figure}

\begin{figure}[!ht]
\begin{center}
   \resizebox{0.99\hsize}{!}{
     \plottwo{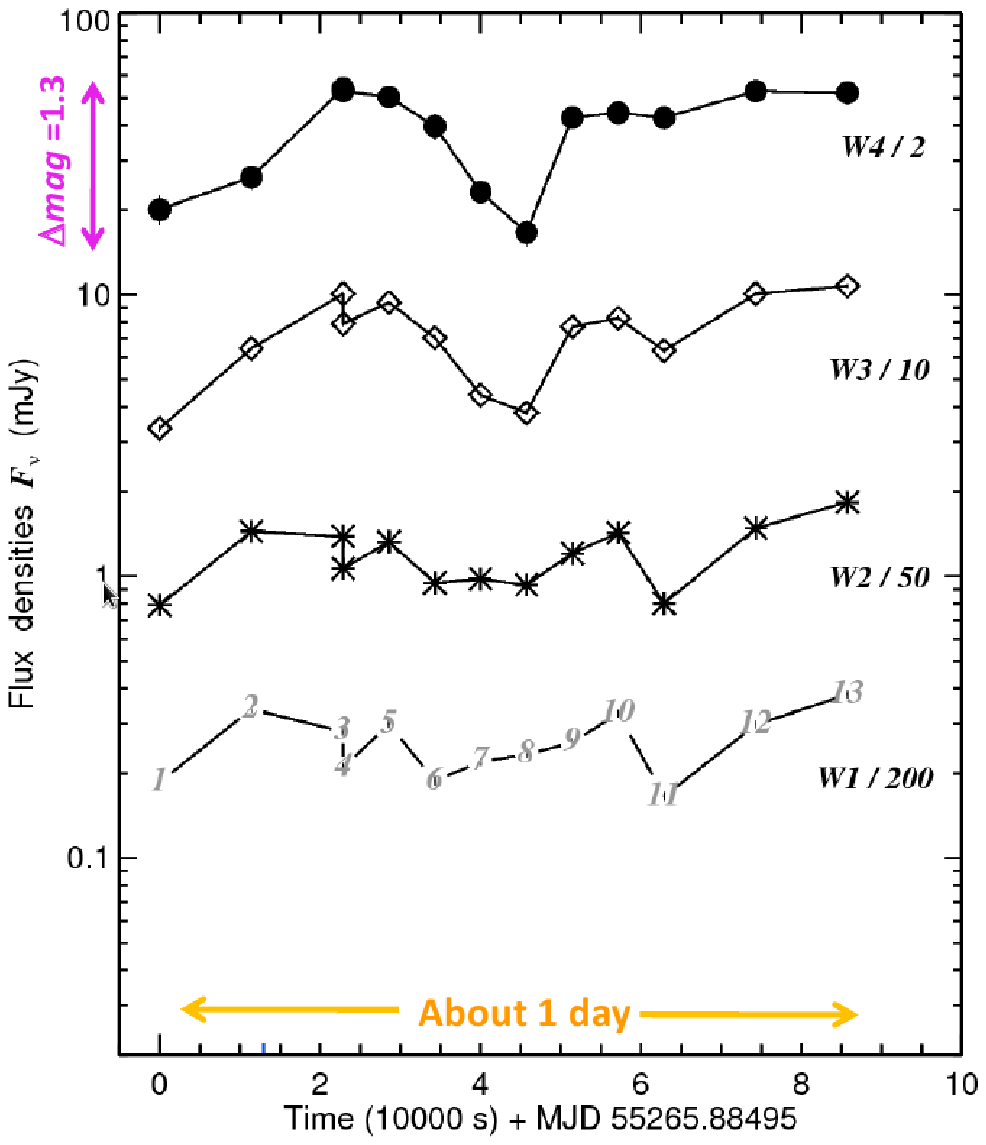}{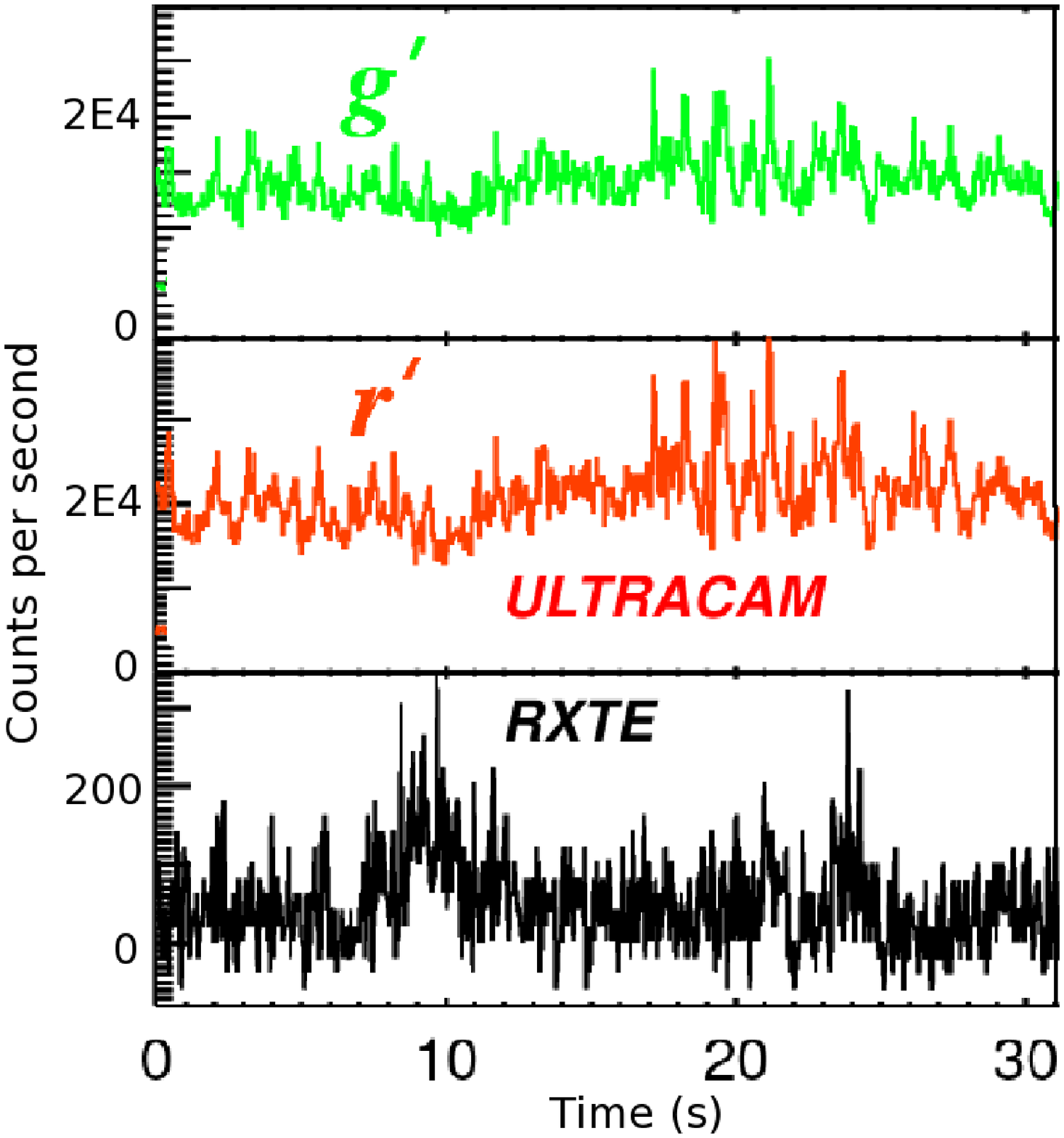}
   }
\end{center}
\caption{
({\em Left}) The first mid-infrared light curve of a stellar-mass BH jet. The data are for GX 339--4 from the WISE mission and span about 24 hours at 13 epochs of observation (each with integration times of $\sim$8--9 s, spaced by a minimum of 11 s (and usually by multiples of $\sim$95 mins, the satellite orbital time). These fully simultaneous observations in four bands W1--4 enabled first direct measurements of $\nu_{\rm break}$ in a black hole binary \citep{g11_wise}. ({\em Right}) A short 30 seconds--long segment of optical and X-ray lightcurves of the same binary in 2010, with rapid sampling of 50 ms \citep{g10}. The optical data ($g'$, $r'$ filters) are from VLT/ULTRACAM; X-rays from {\em RXTE}. As compared to these data, the mid-infrared sampling on the left remains very poor, but this can change with SPICA.
}\label{fig:lc}
\end{figure}

\acknowledgements PG acknowledges support from STFC and an NAOJ visitors grant. He thanks NAOJ/Dr. Masaomi Tanaka for hospitality, and the conference organizers for an interesting meeting. J.M. acknowledges funding from PNHE in France and the french Research National Agency CHAOS project ANR-12-BS05-0009 (http://www. chaos-project.fr). He thanks the IoA Cambridge for hospitality.

\bibliography{pgandhi}

\end{document}